\documentclass{emulateapj}
\usepackage{ulem}
      \def\new#1 {{\bf #1 }}
      \def\cut#1 {\sout{#1} }

\slugcomment{to appear AJ}

\shorttitle{Unveiling a Compact Cluster of Massive and Young stars}
\shortauthors{Zapata et al.}

\begin{document}


\title{Unveiling a Compact Cluster of Massive and Young Stars in IRAS 17233-3606}


\author{Luis A. Zapata\altaffilmark{1}, Silvia Leurini\altaffilmark{2},
        Karl M. Menten\altaffilmark{1}, Peter Schilke\altaffilmark{1},  
      \\Rainer Rolffs\altaffilmark{1} and Carolin Hieret\altaffilmark{1} }

\altaffiltext{1}{Max-Planck-Institut f\"{u}r Radioastronomie, Auf dem H\"ugel 69,
              53121, Bonn, Germany}
\altaffiltext{2}{ESO, Karl Schwarzschild Str. 2, 85748 Garching bei M\"{u}nchen, Germany}


 \begin{abstract}
    We have analyzed sensitive high spatial resolution archival radio continuum data at
    1.3, 2.0, 3.6 and 6.0 cm as well as the H$_2$O maser molecular line data obtained using the
    Very Large Array (VLA) in its hybrid AB configuration toward the high-mass star-forming region
     IRAS 17233-3606 (G351.78-0.54).
    We find nine compact radio sources associated with this region, six of them are new radio detections.
    We discuss the characteristics of these sources based mostly on their spectral indices and find
    that most of them appear to be optically thin or thick ultra- and hyper-compact HII regions ionized
    by B ZAMS stars. Furthermore, in a few cases  the radio emission may arise from optically thick
    dusty disks and/or cores, however more observations at different wavelengths are necessity  
    to firmly confirm their true nature.   
    In addition, we compared our centimeter maps with the mid-infrared images from
    the Spitzer Space Observatory GLIMPSE survey revealing a cluster of young protostars   
    in the region together with multiple collimated outflows some of whom might be related with the compact 
    centimeter objects. Finally, we find that one of these centimeter objects, VLA2d, is well
    centered in an apparent strong and compact north-south bipolar outflow traced by OH masers and
    we therefore suggest that this object maybe is energizing the latter.
 \end{abstract}


\section{Introduction}

Attention to the high-mass star-forming region IRAS 17233-3606 (or G351.78-0.54) 
was first garnered by the detection of very strong maser emission from various species
e.g. hydroxyl (OH) and water (H$_2$O) \citep{CaswellHaynes1980, CaswellHaynes1983}.
At the time of its detection (1980) it possibly contained the strongest
known interstellar hydroxyl maser in the sky, with a peak intensity of 1000 Jy \citep{CaswellHaynes1980}. 
Later on, intense class II methanol (CH$_3$OH) maser emission was detected \citep{Menten1991}.
From the IRAS fluxes and assuming a distance of 2 kpc it has been estimated that IRAS 17233-3606 
has a bolometric luminosity of $\sim$1.5 $\times$ 10$^5$ L$_\odot$ \citep{HughesMacLeod1993}. 
However, more recent measurements with a better estimation of its kinematic near distance (of about 1 kpc) 
suggest a luminosity of 2.5 $\times$ 10$^4$ L$_\odot$ \citep{MacLeodetal1998}. 
We here adopt a distance to IRAS 17233-3606 of 1 kpc, the near kinematic distance
to the object reported by \citet{MacLeodetal1998}. The reason that the far distance appears unlikely
is because the source is located at an angular separation of more than 0.5$^\circ$ from the Galactic plane.

\citet{Argonetal2000}, \citet{Argonetal2002} and \citet{Fishetal2005} using VLA and VLBA radio observations
of all the four hyperfine structure lines of hydroxyl (OH) around 1.6 GHz revealed that the maser spots
clearly show a velocity gradient with a north-south orientation and in a range between -10.4 to 7 km s$^{-1}$.
The masers are centered in the position R.A. = 17$^h$26$^m$42.7$^s $, decl.= -36$^\circ$09$'$17.4$''$ (J2000.0).
\citep{Fishetal2005}. The large broad velocity gradient displayed by the OH masers, several times larger than
in most sources \citep{Argonetal2000} indicates that they could may be tracing a compact bipolar outflow.

\citet{Norrisetal1993} found methanol maser spots coincident with the location of the OH masers
that also appear to display a well-defined north-south velocity gradient.
 \citet{ForsterCaswell1989} and \citet{Forster1990} detected water maser spots
about 3$''$ west of the hydroxyl and methanol maser location.
The water masers cover a much broader velocity range, between $-38$ to +22 km s$^{-1}$ and
seem to be tracing an ``expanding ring'' surrounding possibly a central massive protostar.
 \citet{ForsterCaswell1989} proposed that the compact size ($\sim$ 2$''$), wide velocity spread
in the masers, and a lack of strong continuum emission indicate that this ``ring'' might be produced by
a star in its early expansion phase. However, they also suggested that the geometry could  be due to one
or multiple outflows.

\begin{figure*}[ht]
\centering
\includegraphics[scale=0.32]{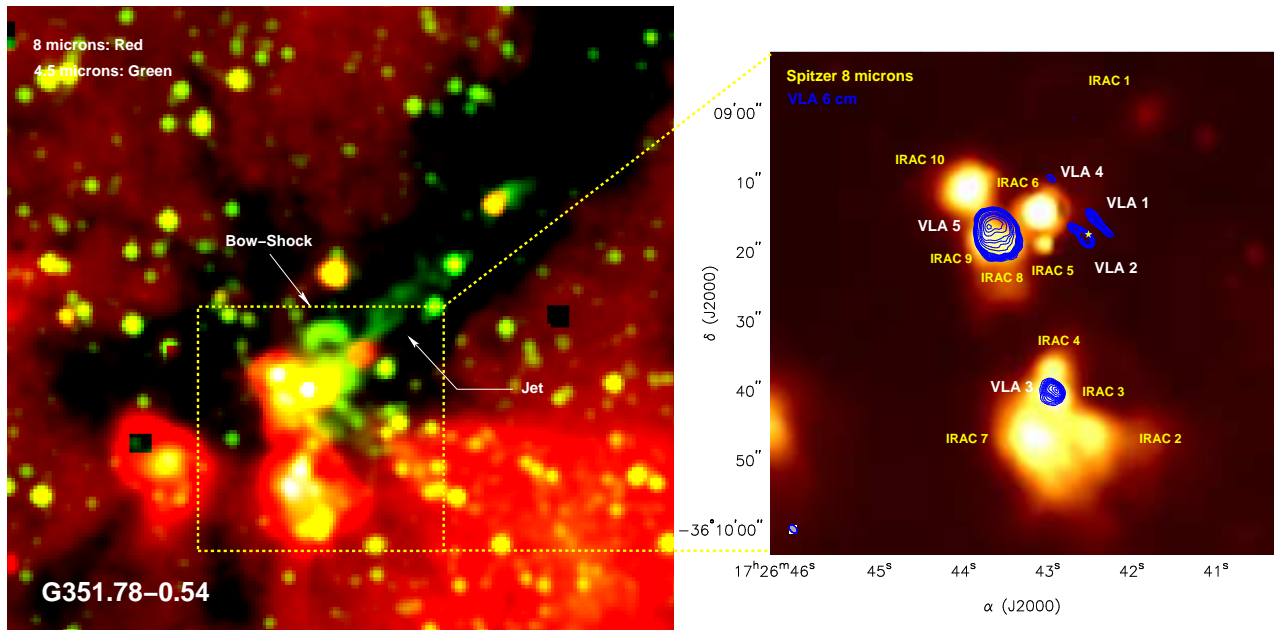}\vspace{0.5cm}
\caption{\scriptsize Spitzer infrared color images of the region IRAS 17233-3606. {\bf Left:}
                     Two image composite with red representing 8.0 $\mu$m and green 4.5 $\mu$m.
                     The angular resolution for the 8.0 $\mu$m and 4.5 $\mu$m images are $\sim$ 1$''$ and 2$''$, respectively.
                     {\bf Right:} 8.0 $\mu$m image of the central
                     part of the IRAS 17233-3606 region overlaid with the 6 cm continuum emission from the VLA.
                     The contours are -3, 3, 4, 5, 6, 7, 8, 9, 10, 12, 14, 16, 18, 20, 40, 60, 80, 100,
                     150, and 200 times 0.1 mJy beam$^{-1}$, the rms noise of the image.
                     The FWHM contour of the synthesized beam is shown in the
                     bottom left corner of the image.
                     The size of the synthesized beam is 1.39$''$ $\times$ 0.88$''$
                     with a position angle (P.A.) of 47.3$^\circ$ E of N.
                     Please note the multiple powerful outflows and jets that emanate
                     from this region traced by the 4.5 $\mu$m band.
                     This IRAC band (4.5 $\mu$m) contains several H$_2$ transitions which trace 
                     shock-excited material \citep{Smithetal2005}.
                     The position of the yellow star in the right image marks the position of the radio binary star reported
                     by  \citet{HughesMacLeod1993} }\vspace{1.0cm}
\label{fig1}
\end{figure*}

A few radio sources have been found in this region, \citet{Haynesetal1979} and \citet{CaswellHaynes1980}
reported for the first time the possible presence of an HII region using observations at 2 cm. 
Subsequent observations at 2 cm by \citet{Fixetal1982} confirmed
this detection and found that this continuum source is extended and bright (310 mJy at 2 cm)
and that it is located almost 10$''$ east of the OH/CH$_3$OH/H$_2$O maser zone.
Later radio observations at 3.6 cm  with a better angular resolution also detected this extended
continuum source and slightly resolved it in a compact HII region with a cometary
morphology \citep{Walshetal1998}.  \citet{Fixetal1982} and \citet{HughesMacLeod1993} report the detection
of a very weak radio continuum source near to the location of the OH/CH$_3$OH/H$_2$O masers.
The observations at 2 cm made by \citet{HughesMacLeod1993} with a resolution
of $0\rlap.''3$ revealed that this object is actually a radio binary and that it is located very close
to the maser positions.

\citet{Fixetal1982} and \citet{Walshetal1999} detected 2.0 $\mu$m infrared sources toward IRAS 17233-3606.
 \citet{Fixetal1982} reported an infrared object close to the location of the masers.
Images by \citet{Walshetal1999} find a source coincident with the OH/CH$_3$OH/H$_2$O masers
and a source coincident with the extended HII region to the east.
Finally, \citet{DeBuizeretal2000} detected a compact and faint infrared object at 10 $\mu$m 
located about 10$''$ east of the position of the masers.
They also detected very diffuse and extended 18 $\mu$m infrared
 emission about 3$''$ northeast of the OH/CH$_3$OH/H$_2$O maser location.

In this paper we present radio continuum observations made with the Very Large Array (VLA) toward the massive
star forming region IRAS 17233-3606. We report the detection of new radio sources associated with this zone
which are distributed among a cluster of strong infrared sources detected by the Spitzer Space Telescope.
We discuss the nature of these radio objects based mainly on their radio spectral indices and
counterparts at other wavelengths. In addition, we present imaging of H$_2$O maser emission associated
with IRAS 17233-3606. 

\begin{deluxetable}{ccccc}[h]
\tablecolumns{5}
\tabletypesize{\footnotesize}
\tablewidth{0pc}
\tablecaption{Summary of the Observations}
\label{table1}
\tablehead{
\colhead{}&
%\multicolumn{2}{c}{Flux Densities} &
\multicolumn{2}{c}{Angular Resolution$^b$} &
\colhead{Rms}&
\colhead{Largest Angular}\\
\cline{2-3}
\colhead{} &
%\colhead{3c286} &
%\colhead{1622-297$^a$} &
\colhead{Beam} &
\colhead{P.A.}&
\colhead{Noise}&
\colhead{Scale}\\
\colhead{Band}&
%\multicolumn{2}{c}{[Jy Beam$^{-1}$]} &
\colhead{[arcsec]}&
\colhead{[degrees]}&
\colhead{[mJy Beam$^{-1}$]}&
\colhead{[arcsec]}
}
\startdata
1.3cm & 0.27$\times$0.25 & 7.1&0.45 & 4\\
2.0cm & 0.51$\times$0.42 & 9.7&0.28 & 7\\
3.6cm & 1.48$\times$0.43 & 40.3&0.06 & 15\\
6.0cm & 1.39$\times$0.88 & 43.7&0.03 & 20\\
Mas.$^a$ & 0.35$\times$0.20 & 34.9&10 & 4
\enddata
\tablecomments{
               (a): The rms-noise is in a velocity channel.\\
               (b): Mayor axis $\times$ minor axis; position angle of mayor axis. }
\end{deluxetable}

\begin{figure*}[!ht]
\centering
\includegraphics[scale=0.6]{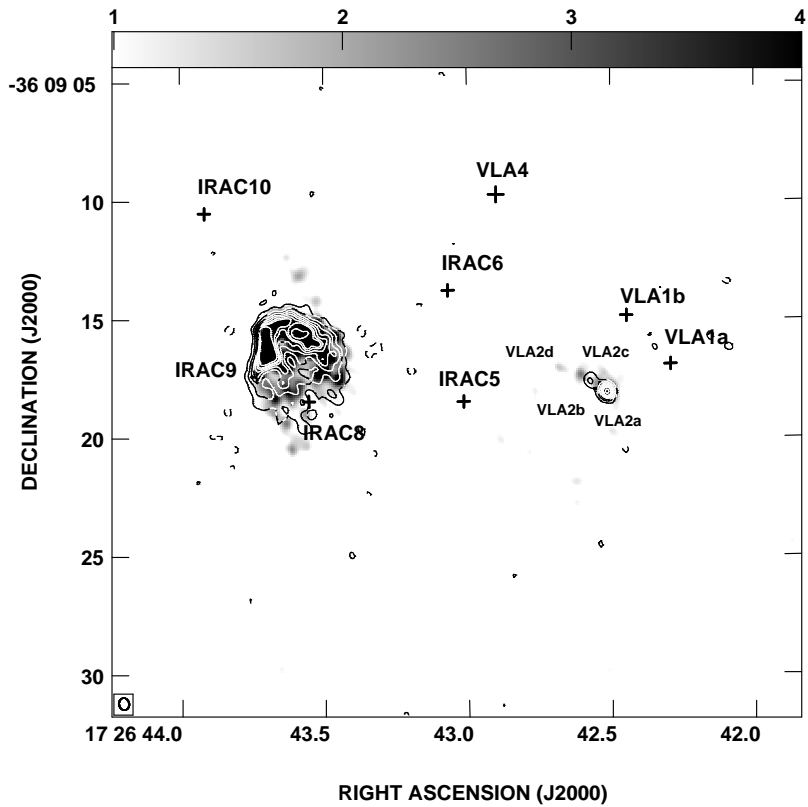}
\caption{\scriptsize VLA 1.3 cm continuum grey scale image overlaid with a blue-contour image at 2.0 cm
                     of a central region of IRAS 17233-3606.
       The contours are -3, 3, 6, 9, 12, 15, 18, 20 and 25 times 0.30 mJy beam$^{-1}$,
       the rms noise of the image. The half-power contour of the synthesized beam is shown in the
       bottom left corner of the image. The size of the synthesized beam is 0.51$''$ $\times$ 0.42$''$
                     with a P.A. of 9.7$^\circ$ E of N. The grey scale bar on the top indicates the 1.3 cm continuum
       emission on mJy beam$^{-1}$. The crosses indicate the positions of the multiple radio and infrared
      sources presented in Figure 1. The 1.3 cm image was convolved to the same angular resolution as the 2.0 cm image, 
using the parameter "UVTAPER" from the task "IMAGER" of AIPS. }\vspace{1.0cm}
\label{fig2}
\end{figure*}

\section{Observations}

From the NRAO\footnote{The National Radio Astronomy Observatory is a facility of the National Science
Foundation operated under cooperative agreement by Associated Universities, Inc.} VLA archival database 
we retrieved continuum data of the
IRAS 17233-3606 region at wavelengths of 1.3, 2.0, 3.6, and 6.0 cm  and spectral line data 
of the 22.2 GHz water maser transition 
(Project number AG502).
The data were taken with the VLA
in its hybrid AB configuration. 
The region was observed between 1997 January 26 and
28 using the 27 antennas of the array.
The phase center was R.A. = 17$^h$26$^m$42.5$^s$, decl.= -36$^\circ$09$'$17.38$''$ (J2000.0).
The amplitude calibrator was 1328+307 (3C286) and the phase calibrator was 1622-297 in all bands.
The flux densities for 1328+307 were 2.5 Jy (1.3 cm), 3.4 Jy (2 cm), 5.2 Jy (3.6 cm) and 7.4 Jy (6.0 cm),
while the bootstrapped flux densities for 1622-297 were 2.3  $\pm$ 0.1 Jy (1.3 cm), 2.66 $\pm$ 0.05 Jy (2.0 cm),
2.5 $\pm$ 0.1 Jy (3.6 cm) and 2.78 $\pm$ 0.05 (6.0cm).  

For the maser observations
the frequency was centered on the H$_2$O $\nu$=1 $J$=6$_{1,6}$-5$_{2,3}$ $F$=6-5 at 22.235077 GHz, while
the correlator was configured in line mode with a band of 64 channels over 6.25 MHz,
which provided 97.6 kHz (1.3 km s$^{-1}$) resolution.

The data were analyzed in the standard manner using the AIPS package of the NRAO. The data were also self-calibrated
in phase and amplitude for each band and for the H$_2$O maser data. Most of the images were made with the
ROBUST parameter of the task IMAGR set to 0, to obtain an optimal compromise between sensitivity and angular resolution.
However, in some cases we used ROBUST set to -5, to get a better angular resolution sacrificing
some sensitivity. The resulting rms noise levels and angular resolution for each band are listed in Table 1.
In this table we also include the largest scale structure to which the array is sensitive.
We assume that the systemic LSR velocity of the molecular cloud associated with IRAS 17233-3606 
is $-4$ km s$^{-1}$. 

\begin{deluxetable}{lcccc}[ht]
\tablecolumns{5}
\centering
\tabletypesize{\footnotesize}
\tablewidth{0pc}
\tablecaption{Parameters of the 8 microns Infrared Sources}
\label{table2}
\centering
\tablehead{
\colhead{IRAC} &
\multicolumn{2}{c}{Position} &
%\colhead{Density Flux} &
\colhead{Size$^a$} &
\colhead{P.A.$^b$}\\
\cline{2-3}
\colhead{ }&
\colhead{$\alpha_{2000}$}&
\colhead{$\delta_{2000}$}&
%\colhead{[mJy beam$^{-1}$]}&
\colhead{[arcsec]}&
\colhead{[deg.]}
}
\startdata
1 & 17 26 42.845 & -36 08 35.23 & 3.3 $\times$ 2.9 & 16$^\circ$\\
2 & 17 26 42.529 & -36 09 46.16 & 9.2 $\times$ 7.3 & 49$^\circ$\\
3 & 17 26 42.946 & -36 09 40.65 & 6.1 $\times$ 5.2 & 22$^\circ$\\
4 & 17 26 42.950 & -36 09 37.69 & 6.3 $\times$ 4.6 & 164$^\circ$\\
5 & 17 26 43.040 & -36 09 18.38 & 3.6 $\times$ 3.1 & 155$^\circ$\\
6 & 17 26 43.075 & -36 09 14.02 & 3.2 $\times$ 3.2 & 82$^\circ$\\
7 & 17 26 43.082 & -36 09 46.48 & 12.5$\times$ 9.8 & 27$^\circ$ \\
8 & 17 26 43.555 & -36 09 18.48 & 7.4 $\times$ 5.8 & 69$^\circ$  \\
9 & 17 26 43.708 & -36 09 17.94 & 5.8 $\times$ 5.2 & 14$^\circ$\\
10 & 17 26 43.920& -36 09 10.84 & 6.0 $\times$ 5.7 & 12$^\circ$
\enddata
\tablecomments{  Units of right ascension are hours, minutes, and seconds, and units of declination are degrees,
                 arcminutes, and arcseconds.\\
                 (a): Mayor axis $\times$ minor axis
                 (b): Position angle of mayor axis.}

\end{deluxetable}

\begin{deluxetable*}{lcccccc}[h]
\tablecolumns{6}
\tabletypesize{\footnotesize}
\tablewidth{0pc}
\tablecaption{Parameters and Tentative Nature of the Centimeter Sources}
\label{table2}
\centering
\tablehead{
\colhead{Wavelength} &
\multicolumn{2}{c}{Position} &
\colhead{Flux Density} &
\colhead{Deconvolved Size$^a$} &
\colhead{P.A.$^b$}\\
\cline{2-3}
\colhead{cm }&
\colhead{$\alpha_{2000}$}&
\colhead{$\delta_{2000}$}&
\colhead{[mJy beam$^{-1}$]}&
\colhead{[arcsec]}&
\colhead{[deg.]}
}
\startdata

& & & {\bf VLA1a } & &\\
\\

6.0  &  17 26 42.446 & -36 09 14.89   &  7.3$\pm$0.3    &  5.4$\pm$0.1 $\times$ $\leq$ 1  & 42$\pm$1       \\
3.6  &  17 26 42.378 & -36 09 15.80   & 4.8$\pm$0.3     &  4.4$\pm$0.5 $\times$ 0.24$\pm$0.05  & 41$\pm$1       \\
2.0  &   -     &    -        &  $\leq$ 1 &   --      &  --       \\
1.3  &   -     &    -        &  $\leq$ 2&   --      &  --       \\

\\
& & & {\bf VLA1b} & &\\
\\

6.0 &  17 26 42.543 & -36 09 17.92   & 11.4$\pm$0.1    &  1.0$\pm$0.1 $\times$ 0.31$\pm$0.05  & 40$\pm$1  \\
3.6 &  17 26 42.418 & -36 09 15.24   & 3.2$\pm$0.3     &  1.8$\pm$0.5 $\times$ 0.34$\pm$0.05  & 41$\pm$1        \\
2.0 &   -     &    -        &  $\leq$ 1 &   --      &  --       \\
1.3 &   -     &    -        &  $\leq$ 2&   --      &  --       \\

\\
& & & {\bf VLA2a } & &\\
\\

6.0 &  17 26 42.540 & -36 09 17.95   &  4.0$\pm$0.3    &  $\leq$ 2                   & -- \\
3.6 &  17 26 42.529 & -36 09 18.14   & 12.3$\pm$0.3    &  1.1$\pm$0.5 $\times$ 0.35$\pm$0.05  & 43$\pm$1  \\
2.0 & 17 26 42.514 & -36 09 18.10 &  10.1$\pm$0.5   & 0.28$\pm$0.05 $\times$ 0.19$\pm$0.05 & 59$\pm$41  \\
1.3 & 17 26 42.508 & -36 09 18.04    &  11.8$\pm$0.7   & 0.10$\pm$0.05 $\times$ $\leq$ 0.05 & --   \\

\\
& & & {\bf VLA2b } & &\\
\\

6.0 &  17 26 42.540 & -36 09 17.95   &  4.0$\pm$0.3    &  $\leq$ 2                   & -- \\
3.6 &  17 26 42.526 & -36 09 18.18   &  2.7$\pm$0.3    &   $\leq$ 2                    & -- \\
2.0 & 17 26 42.518 & -36 09 18.10 &  3.4$\pm$0.5    & 0.53$\pm$0.05 $\times$ 0.25$\pm$0.05& 49$\pm$18 \\
1.3 & 17 26 42.547 & -36 09 17.67    &  3.0$\pm$0.5   & 0.34$\pm$0.05 $\times$ 0.10$\pm$0.03& 153$\pm$63 \\

\\
& & & {\bf VLA2c } & &\\
\\

6.0 &   -           &   -            &   $\leq$ 0.1               &                       --            & -- \\
3.6 &   -      &   -        &   $\leq$ 0.2               &                       --            & -- \\
2.0 &   -     &   -         &   $\leq$ 1               &                       --            & -- \\
1.3 & 17 26 42.609 & -36 09 17.35    &  3.7$\pm$0.8   & 0.29$\pm$0.05 $\times$ $\leq$ 0.1 & 34$\pm$11 \\

\\
& & & {\bf VLA2d } & &\\
\\

6.0 &   -           &   -            &   $\leq$ 0.1               &                       --            & --\\
3.6 &   -     &   -         &   $\leq$ 0.2               &                       --            & --\\
2.0 &   -     &   -         &   $\leq$ 1               &                       --            & --\\
1.3 & 17 26 42.686 & -36 09 17.05    &  2.7$\pm$0.7   & 0.18$\pm$0.05 $\times$ $\leq$ 0.1 & 46$\pm$24 \\

\\
& & & {\bf VLA3 } & &\\
\\

6.0  &  17 26 42.934 & -36 09 39.87   & 11.2$\pm$0.3    &  2.2$\pm$0.1 $\times$ 2.0$\pm$0.1 & 53$\pm$13      \\
3.6  &  17 26 42.922 & -36 09 39.97   & 5.3$\pm$0.3     &  1.9$\pm$0.5 $\times$ 1.0$\pm$0.1 & 42$\pm$6        \\
2.0  &    -    &    -        & $\leq$ 1 &   --      &   --       \\
1.3  &    -    &    -        & $\leq$ 2 &   --      &   --       & \\

\\
& & & {\bf VLA4 } & &\\
\\

6.0  &  17 26 42.940 & -36 09 09.31   & 9.8$\pm$0.1     &  1.2$\pm$0.2 $\times$ $\leq$ 1 & 65$\pm$5.5        \\
3.6  &  17 26 42.933 & -36 09 09.48   & 0.8$\pm$0.05     &  0.90$\pm$0.05 $\times$ 0.30$\pm$0.05   & 38$\pm$17      \\
2.0  &    -    &    -        & $\leq$ 1 &   --      &   --       \\
1.3  &    -    &    -        & $\leq$ 2 &   --      &   --       & \\

\\
& & & {\bf VLA5 } & &\\
\\

6.0  &  17 26 43.621 & -36 09 16.31   & 330$\pm$5       &  3.1$\pm$0.1 $\times$ 2.9$\pm$0.1   &  112$\pm$1   \\
3.6  &  17 26 43.691 & -36 09 16.46  & 252$\pm$5   &    3.1$\pm$0.1 $\times$ 2.9$\pm$0.1      &  103$\pm$2  \\
2.0  &  17 26 43.699 & -36 09 16.37 &  211$\pm$5&   3.1$\pm$0.1 $\times$ 2.2$\pm$0.1 & 105$\pm$3  \\
1.3  & 17 26 43.696 &   -36 09 16.27 &  300$\pm$5 &   $\leq$ 3      & --  \\
&   -------------------- & & & & \\
&  {\bf Possible Nature} & & & & \\

VLA1a &   ?  &     &    &       &  \\
VLA1b &   ? &     &    &        &      \\
VLA2a &    HCHII region? &     &    &        &\\
VLA2b &    HCHII region? &     &    &         &  \\
VLA2c &    HCHII region or Dusty Disk? &    &     &        &\\
VLA2d &    HCHII region or Dusty Disk?  &    &     &        & \\
VLA3  &  Optically Thin UCHII region  &     &     &       &      \\
VLA4  &  Optically Thin UCHII region?  &     &     &       &     \\
VLA5  &  Optically Thin UCHII region &     &     &       &   \\
\\
\enddata
\tablecomments{Units of right ascension are hours, minutes, and seconds, and units of declination are degrees,
arcminutes, and arcseconds.\\
(a): Mayor axis $\times$ minor axis\\
(b): Position angle of mayor axis.}
\end{deluxetable*}

\section{Results and Discussion}

In a region of about 1$'$ $\times$ 1$'$, we detected a total of 5 compact  sources in one or more
of the individual bands, as well as 16 water maser spots;  see Figures 1, 2, and 3.  However, two of the radio
sources, VLA1 and VLA2, were resolved into 2 (VLA1a,b) and 4 (VLA2a,b,c,d) compact sources,
respectively, see Figures 2 and 3.
  The observed parameters were determined from a linearized
least-square fit to an elliptical Gaussian using the task IMFIT of AIPS.

Infrared two color band (8 and 4.5 $\mu$m) images from the Spitzer Space Observatory's GLIMPSE survey \citep{Benjaminetal2003}
overlaid  with our 6 cm radio image of the IRAS 17233-3606 (both with similar angular reslotions) are presented in Figure 1. 
In this image a cluster of 10 bright  8 $\mu$m infrared
sources is detected. We give the positions and sizes of the sources in Table 2 obtained  
using the task IMFIT of AIPS. These infrared sources are possibly associated with the objects detected by \citet{Fixetal1982}, 
\citet{Walshetal1999} and \citet{DeBuizeretal2000} at IR wavelengths.
In a similar manner we give the observed parameters of the radio sources and the masers
in Tables 3 and 4.

As it can be seen, some of these infrared sources appear to be related with the centimeter objects,
however, the nature of these infrared objects will not be discussed in this article, they are presented here
only for a reference.

In what follows, we discuss separately each of the 9 radio sources detected in this region.
Our discussion of the spectral indices is based on the assumption that the flux densities did not change
between the 1997 January 26 and 28 observations. Furthermore, due to the spectral index measurements presented here can
be very sensitive to the largest angular size and the angular resolution of the observations, 
matching-beam observations of the sources at several frequencies are thus needed to discuss 
them in a very reliable way.

\begin{figure*}[!ht]
\centering
\includegraphics[scale=0.4]{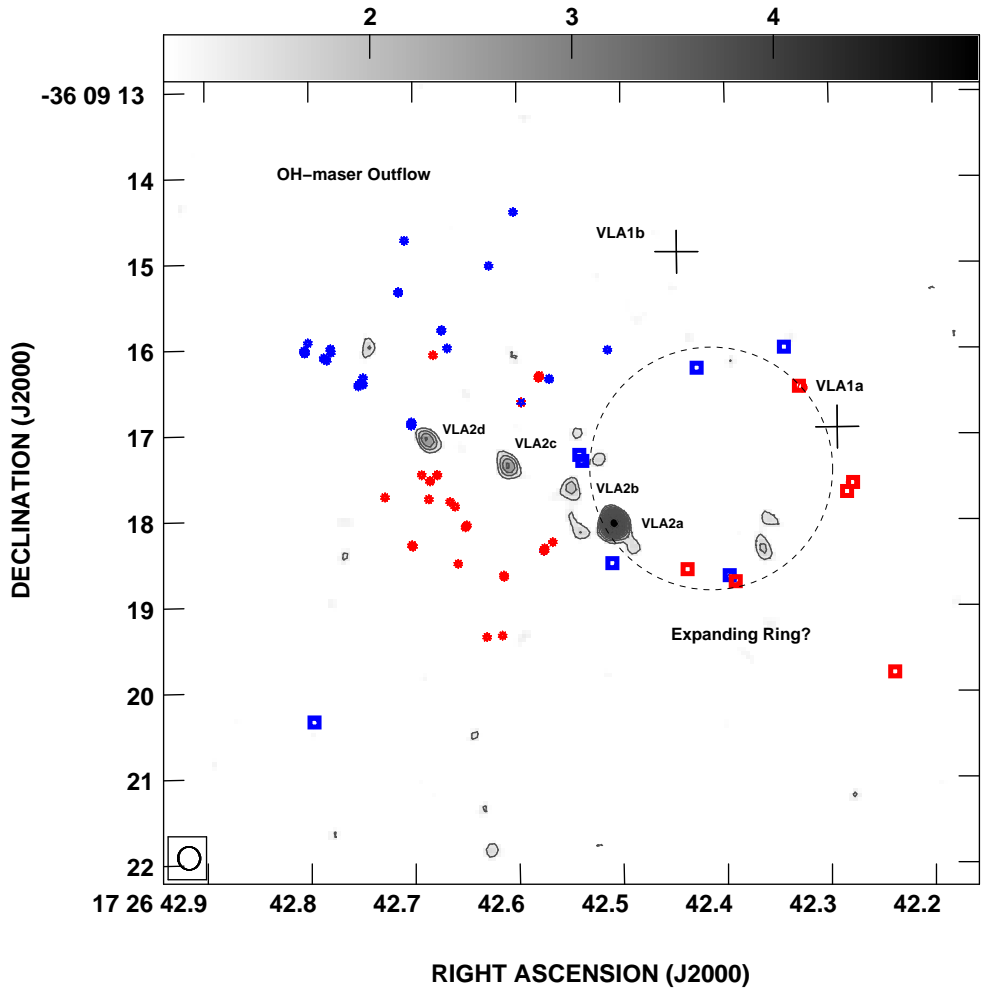}
\caption{\scriptsize VLA 1.3 cm continuum image of the maser zone.
                     The contours are -3, 3, 4, 5, 6, 7, 8, 9, 10, 11, 12, 13, 14 and 15 times 0.43 mJy beam$^{-1}$,
                     the rms noise of the image. The half-power contour of the synthetized beam is shown in the
                     bottom left corner of the image. The size of the beam is 0.27$''$ $\times$ 0.25$''$
                     with a P.A. of 7.1$^\circ$. The grey scale bar on the top indicates the 1.3 cm continuum
                     emission on mJy beam$^{-1}$. The blue and red dots indicate the positions of the blue- and
                     red-shifted OH maser spots, respectively  (Fish et al. 2005). Note that these masers
                     appear to be tracing the innermost part of  a north-south compact outflow that shows a moderate
                     velocity gradient, going from -10.4 to 7 km s$^{-1}$ (Fish et al. 2005) and that seems to emanate
                     from the radio source VLA2d.
                     The blue and red open squares (about 3$''$ west from the OH center maser) indicate the position
                     of the water masers presented on this article (see Table 4) and that seem to be tracing
                     an ``expanding ring'' (Foster 1990). The two black crosses mark the position
                     of the objects VLA1a and b.  }
\label{fig3}\vspace{1.5cm}
\end{figure*}

\subsection{Classical Ultra-compact HII Regions}

\subsubsection{VLA5}

As already mentioned in the introduction this source was discovered and reported as a faint HII region
for the first time by \citet{Haynesetal1979} and \citet{CaswellHaynes1980}. Later
observations with a better signal to noise and angular resolution by \citet{Fixetal1982},
\citet{Walshetal1998}, and \citet{Argonetal2000} confirm its detection. However, this source is not
related with the center of the OH, CH$_3$OH and H$_2$O masers ({\it the maser zone}), but rather it is offset by about
10$''$ east , see Figures \ref{fig2}, \ref{fig3}, and the images from \citet{Walshetal1998}
and \citet{Argonetal2000}.

Figure 2 shows the 1.3 and 2 cm continuum emission from the central portion of the IRAS 17233-3606 region,
the HII region and the radio binary system reported by \citet{HughesMacLeod1993} are
shown in this image.  The HII region is resolved and shows a very clumpy cometary morphology.
We give its deconvolved dimensions and flux densities for all bands in Table 3.

\begin{figure*}[!ht]
\centering
\includegraphics[scale=0.35]{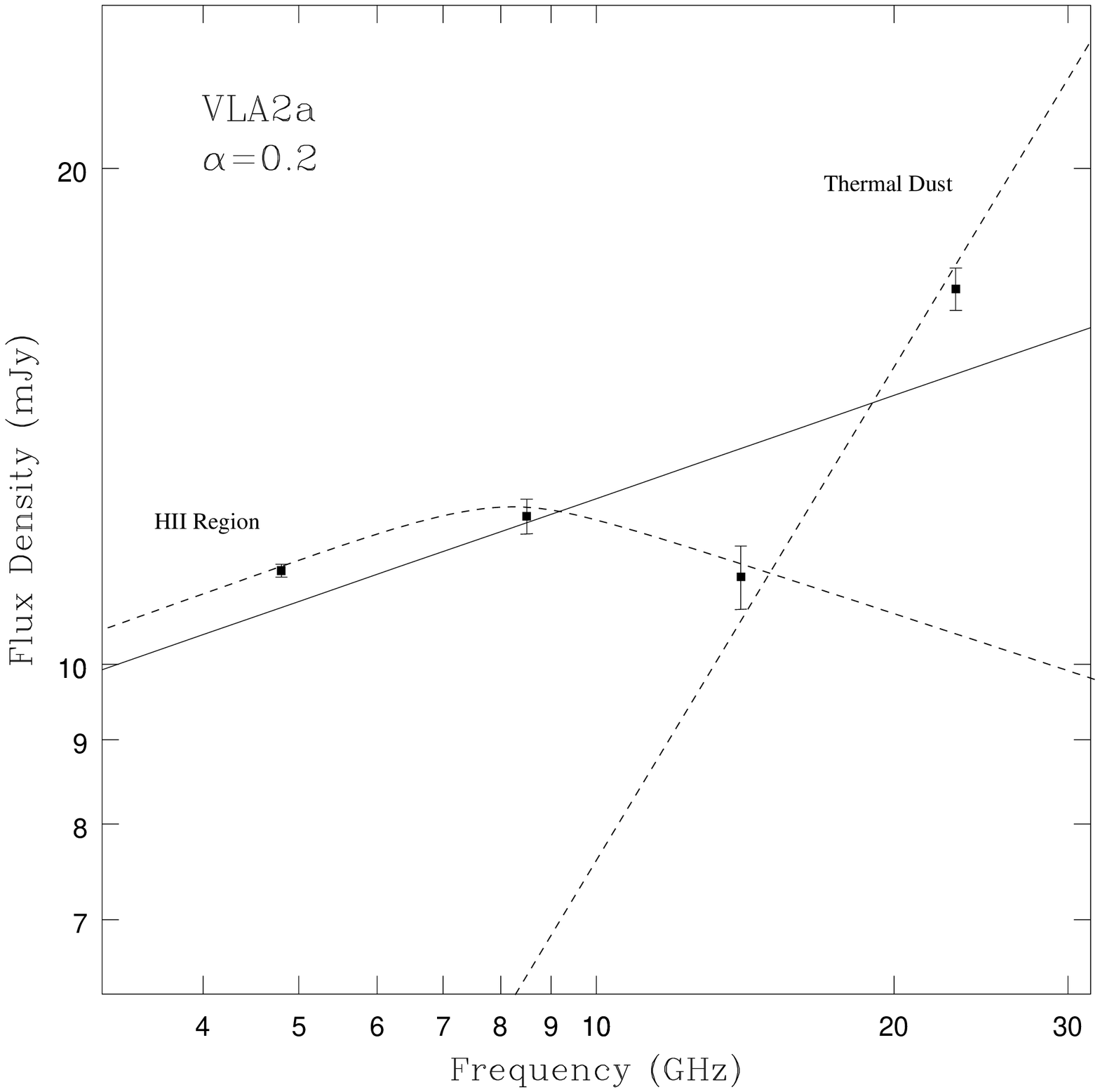}
\includegraphics[scale=0.35]{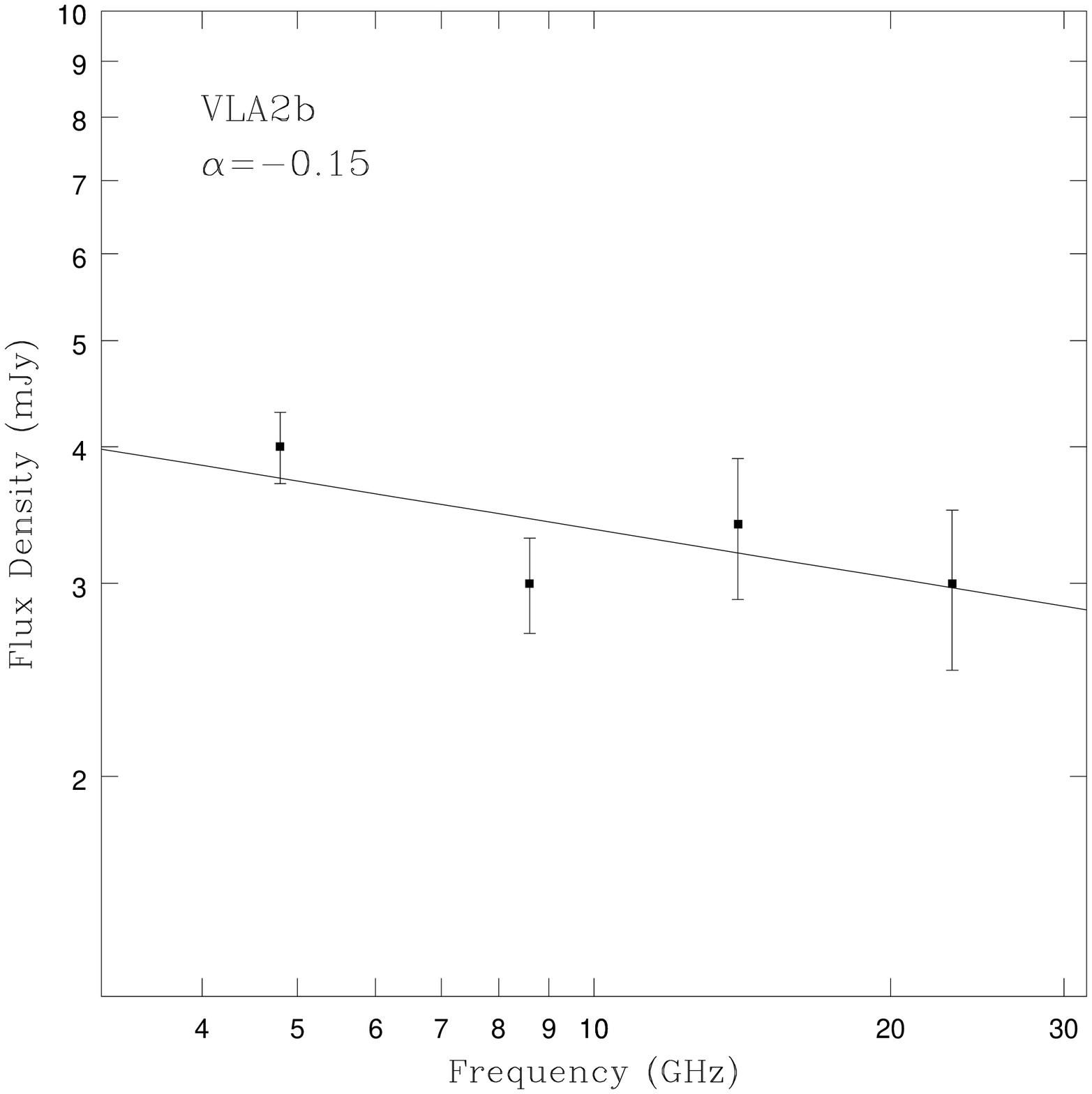}\\
\includegraphics[scale=0.35]{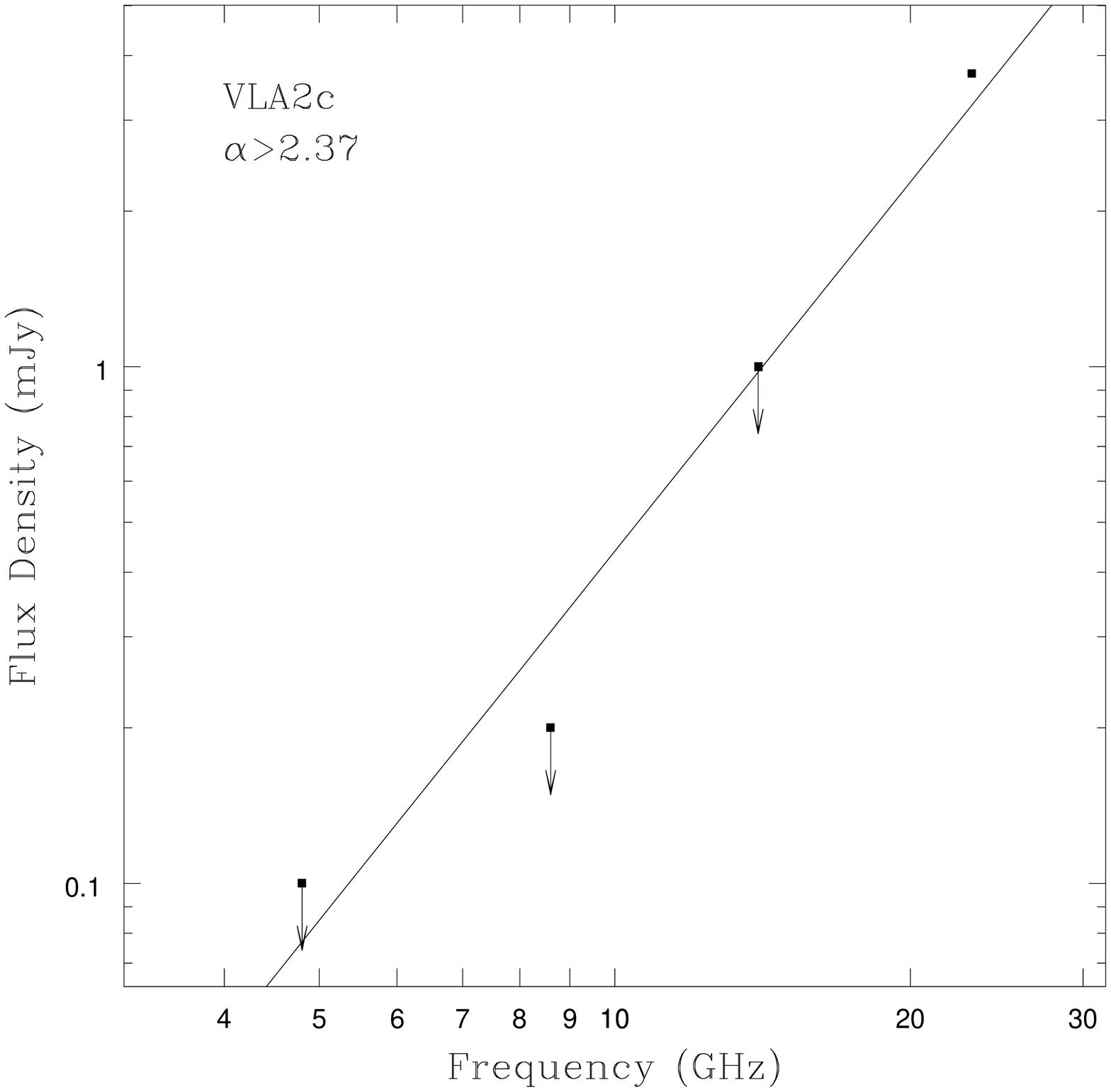}
\includegraphics[scale=0.35]{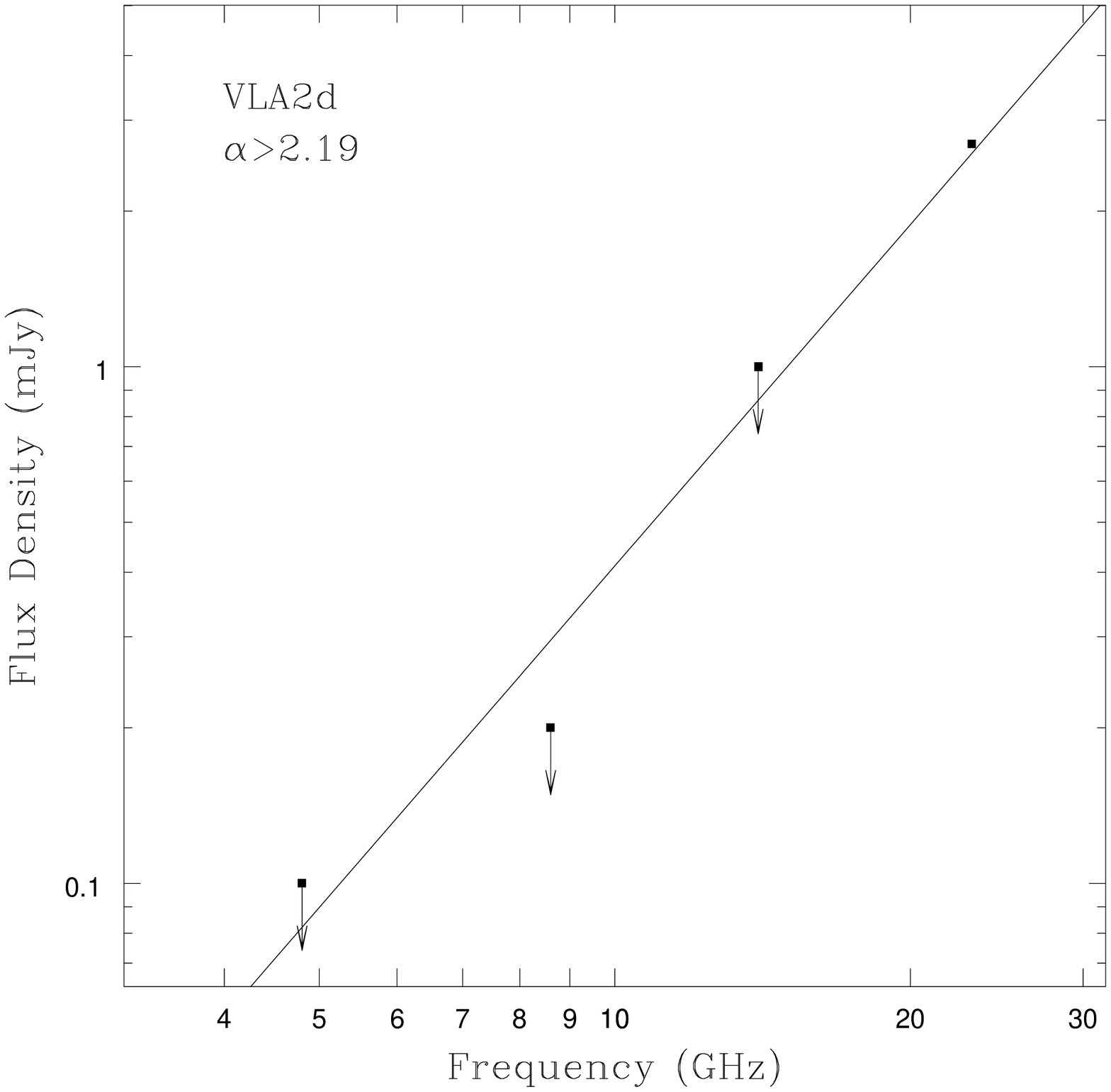}
\caption{\scriptsize Spectral energy distributions of the radio sources VLA2a-d from about 3 to 30 GHz (or 10 to 1 cm).
The squares are detections.
The squares with arrows are upper limits (4$\sigma$). In cases where the error bars were smaller than 
the squares they are not presented.  The line is a least-squares power-law fit
(of the form S$_\nu$ $\propto$ $\nu^\alpha$) to each spectrum. For VLA2a and b the flux densities at 14 and 23 GHz (2.1 and 1.3 cm)
were obtained from maps convolved at a resolution of 1$''$. 
The spectrum for VLA2a can be also modeled by a 
two-component spectrum from a classical HII region plus thermal dust emission.}
\label{fig4}
\end{figure*}

The total 2 cm flux densities for this source reported by \citet{Fixetal1982} and us (see Table 3)
are 310 and 330 mJy, respectively. We consider the agreement very good. Furthermore, the 3.6 cm peak
flux densities reported by \citet{Walshetal1998}, and \citet{Argonetal2000} also seem to be consistent
with our measurements.

Taking the flux densities at 2 and 1.3 cm from Table 3, which both have a similar angular resolution (see Table 1),
we estimated a very flat spectral index for this object, equal to 0.06, 
indicating thus that the emission might come from an optically thin UCHII region.

Following the derivation of \citet{vanandmentenetal2005}, and assuming that the free-free emission at 2 cm is 
optically thin
(which must be the case for the above calculation),
that the ionized gas has a temperature of 10$^4$ K, that this source is located a distance of 1 kpc, and 
taken the values of the number of ionizing photons corresponding to the spectral type for a young star
from \citet{Panagiaetal1973}, we estimate that possibly the UCHII region is ionized by a B0.5 ZAMS star. 
This is again in very good 
agreement with the values and spectral types reported in \citet{Walshetal1998}
and \citet{HughesMacLeod1993} .

From Figure 1 we can see that this HII region is coincident with the compact infrared sources,
IRAC 8 and 9, suggesting that maybe they are the stars responsible for its ionization. 
However, both infrared sources seem to lie south and east of the UCHII region.

\subsubsection{VLA3}

This source is first detected here and is imaged at 6 cm in Figure \ref{fig1}.
We note that VLA 3 at this wavelength is resolved and turns out to be 
an UCHII region with a shell-like morphology.
We give its deconvolved dimensions and flux densities for all bands in Table 3.

In a similar way than VLA 5, taking the flux densities at 2 and 1.3 cm from Table 3 for VLA3 (both measurements with 
similar angular resolution), we estimated a spectral index for this object of $\leq$ 1.1, consistent with
the emission also coming from an optically thin HII region.

In addition, assuming again that the free-free emission at
2 cm is optically thin, that the ionized gas has a temperature of 10$^4$ K, and that
this source is located  a distance of 1 kpc, we estimate that this HII
region is possibly ionized by a B2 type ZAMS star.

From Figure 1, we also see that this UCHII region is coincident with one
compact infrared source, IRAC 3.

\subsection{On the Nature of VLA1 and VLA4}

These sources were only detected in our more sensitive maps, at 6 and 3.6 cm, see Figure \ref{fig1}
for an image at 6 cm. VLA4 is very compact and faint, while VLA1 is quite elongated in
northeast-southwest orientation at 6 cm. Furthermore, VLA1 is resolved into two sources (VLA1a,b)
in our 3.6 cm map with a better angular resolution. We give their flux densities and deconvolved
sizes at this wavelength in Table 3.

From the values of the flux densities at 6 and 3.6 cm presented in Table 3, which also both have a similar angular resolution 
(see Table 1), we estimate a spectral
index for VLA4 of $\leq$ 1.0, and for VLA1a and b of $\leq$ -0.8 suggesting that possibly
the emission arises from optically thin HII regions or from ionized stellar winds (the latter case is more likely for VLA4).
The radio emission from VLA1a and b also might be associated with synchrotron emission.
This kind of emission is believed to trace gyrosynchrotron emission from active magnetospheres of young, low-mass stars.
However, similar negative spectral indices also have found in sources associated with high- and low-mass stars and are believed 
to be produced in strong shocks, see \citet{Garayetal1996}.

However, these flattened spectral indices could also be due to missing flux density in the higher angular resolution and 
higher frequency observations, see Table 1 where we include the largest scale structure to which the array is sensitive.

None of these radio sources have a counterpart at infrared wavelengths (8.0 and 4.5 $\mu m$), see Figure 1.
\vspace{-0.5cm}
\begin{deluxetable*}{lcccccccc}[!h]
\tabletypesize{\scriptsize}
\tablecolumns{9}
\tablewidth{0pc}
\tablecaption{Parameters of the Water Masers Spots}
\label{table3}
\centering
\tablehead{
\colhead{Maser} &
\multicolumn{2}{c}{Position} &
\colhead{Central Vel.} &
\colhead{Deconvolved Size$^a$} &
\colhead{P.A.$^b$} &
\colhead{$\int$ S$_\nu$ dv}\\
\cline{2-3}
\colhead{Feature}&
\colhead{$\alpha_{2000}$}&
\colhead{$\delta_{2000}$}&
\colhead{[Km s$^{-1}$]}&
\colhead{[arcsec]}&
\colhead{[deg.]}&
\colhead{[Jy beam$^{-1}$ km s$^{-1}$]}
}
\startdata

1  & 17 26 42.155 & -36 09 11.76 & -1.0 & 0.14$\pm$0.03 $\times$ $\leq$ 0.2        &  173$\pm$80    &     2.1$\pm$0.2 \\
2  & 17 26 42.239 & -36 09 19.77 & -1.0 & 0.40$\pm$0.01 $\times$ $\leq$ 0.2        &   30$\pm$1     &    11.0$\pm$0.2 \\
3  & 17 26 42.280 & -36 09 17.57 &  0.5 & 0.49$\pm$0.02 $\times$ 0.16$\pm$0.01     &  159.6$\pm$0.2 &   181.2$\pm$0.3 \\
4  & 17 26 42.284 & -36 09 17.68 & -0.5 & 0.33$\pm$0.01 $\times$ $\leq$ 0.3        &  140.8$\pm$0.3 &   154.3$\pm$0.3 \\
5  & 17 26 42.326 & -36 09 16.46 & -1.5 & 0.06$\pm$0.03 $\times$ 0.01$\pm$0.06     &  159$\pm$2     &    89.8$\pm$0.2 \\
6  & 17 26 42.343 & -36 09 15.98 & 8/1.5/-19.5 & 0.28$\pm$0.01 $\times$ 0.02$\pm$0.01 &   26$\pm$2     &    13.6$\pm$0.2 \\
7  & 17 26 42.392 & -36 09 18.69 & -2.0 & $\leq$ 0.5 $\times$ $\leq$ 0.5          &   --           &    23.0$\pm$0.5 \\
8  & 17 26 42.395 & -36 09 18.67 & 22.5/0.5/-10  & 0.330$\pm$0.002 $\times$ 0.052$\pm$0.007&63.4$\pm$0.5 &    64.3$\pm$0.6 \\
9  & 17 26 42.428 & -36 09 16.21 & -14.0 & 0.05$\pm$0.01 $\times$ $\leq$ 0.02   &   35$\pm$5     &    58.2$\pm$0.2 \\
10 & 17 26 42.438 & -36 09 18.57 & -2.5  & 0.37$\pm$0.01 $\times$ 0.05$\pm$0.01 &   90.4$\pm$0.3 &   138.3$\pm$0.3 \\
11 & 17 26 42.511 & -36 09 18.49 & -4.0  & 0.13$\pm$0.01 $\times$ 0.08$\pm$0.01 &   27$\pm$2     &    34.0$\pm$0.2 \\
12 & 17 26 42.539 & -36 09 17.28 & -3.5  & 0.13$\pm$0.01 $\times$ 0.02$\pm$0.01 &  168.6$\pm$0.5 &   119.0$\pm$0.2 \\
13 & 17 26 42.538 & -36 09 17.31 & -8.5/-13.5 & $\leq$ 0.5 $\times$ $\leq$ 0.5  &   --           &    21.0$\pm$0.5 \\
14 & 17 26 42.798 & -36 09 20.35 & -11.5 & $\leq$ 0.3 $\times$ $\leq$ 0.1           &   --           &     3.7$\pm$0.5 \\
15 & 17 26 42.902 & -36 09 09.06 & -5.5  & $\leq$ 0.3 $\times$ $\leq$ 0.2           &   --           &     5.9$\pm$0.5 \\
16 & 17 26 42.955 & -36 09 09.15 & -4.5  & $\leq$ 0.3 $\times$ $\leq$ 0.2           &   --           &     3.3$\pm$0.5 \\
\enddata
\tablecomments{Units of right ascension are hours, minutes, and seconds, and units of declination are degrees, arcminutes,
and arcseconds.\\
(a): Mayor axis $\times$ minor axis\\
(b): Position angle of mayor axis.}
\end{deluxetable*}

\subsection{Radio Objects Associated with The Maser Zone}

In Figure 3 we show the 1.3 cm radio emission together with
the position of the OH, CH$_3$OH and H$_2$O masers spots
reported by \citet{Fishetal2005}, \citet{Forster1990} and \citet{Caswelletal1997}.
We found six compact radio sources in this region, VLA2a,b,c,d (VLA2a,b were detected for the
first time by \citet{HughesMacLeod1993} at 6 cm) and VLA1a,b.
In Figure 2 we also show the radio sources detected by
\citet{HughesMacLeod1993}, VLA2a and VLA2b, they appear to form a binary system with
a spatial separation of about 800 AU. However, the new observations show that possibly these sources
are part of a multiple system. The sources VLA2c and d are only
detected in our 1.3 cm images.

Figure 4 shows the SEDs for the centimeter compact sources VLA2a,b,c, and d.
The flux densities values for the four wavelenghts were obtained from Table 3,
however the values at 2 and 1.3 cm were obtained using a convolved map at a resolution of 1$''$, {\it i.e.} 
with a angular resolution similar to those at 3.6 and 6 cm (see Table 1). 
We convolved the maps using the parameter "UVTAPER" from 
the task "IMAGER" of AIPS. This parameter set a Gaussian taper 
to weight down long baseline data points. 

VLA2c and d have steeper spectral indices, $\geq$ 2.2 consistent 
with either optically thick hyper-compact HII regions or with optically thick thermal dust emission
from massive dusty cores and/or disks. If the emission is arising from 
thermal dust, the spatial sizes (about 300 AU) of these
sources indicate that they are more likely to be compact circumstellar disks rather than dusty cores.
VLA2b shows a negative spectral index $\leq$ -0.15, possibly associated with synchrotron emission.
The spectrum of the source VLA2a also can be modeled by a 
two-component spectrum where for the centimeter wavelengths is dominated for a classical HII region,
while on the millimeter wavelengths is dominated by a component that rises rapidly with frequency. 
This component is likely to be associated with dust emission from a core or disk, see Figure 4.

From Figure 3 we note that the radio source VLA2d is well centered in an apparent compact bipolar north-south outflow
traced by the OH maser spots.  
This again suggests that VLA2d maybe is a compact circumstellar disk rather than an hyper-compact HII region.
We also note that there is a strong bow-shock about 40$''$ north
of this compact outflow  and which is quite well aligned to this compact outflow traced by OH masers, see Figure 1.
Possibly this bow-shock is part of an older ejection of this compact outflow.
If this is the case, the outflow should have a dynamical age of  the order of 10$^4$ years, assuming a velocity of 10 km s$^{-1}$.
Furthermore, \citet{Leurinietal2008} found a strong north-south $^{12}$CO outflow using the Atacama Pathfinder EXperiment (APEX) 
radiotelescope and that is emanating from the maser zone, it is likely that this molecular compact outflow forms part of 
this bow-shock and the very compact outflow traced by the OH maser.
 
Finally, there is another highly collimated northwest-southeast jet that seems to be emanating
from this region (see Figure 1), possibly also associated with these compact radio objects.

We do not find any 8 $\mu$m infrared source associated with these four radio objects, indicating 
that maybe they are highly embedded in the molecular cloud. 
The extended 4.5 $\mu$m
source associated with this zone (see Figure 1) seems to be more likely associated with shocked gas possibly from the
multiple outflows originating here

\subsection{An Expanding Ring?}

As first suggested by \citet{ForsterCaswell1989} and as can be observed in Figure 3 
the water maser spots located to the east of the OH masers
appear to be tracing a ``ring''  centered on the position R.A. = 17$^h$26$^m$42.4$^s$
decl.= -36$^\circ$09$'$17.5$''$ (J2000.0) with a radius of approximately 1.5$''$.
The maser spots associated with the ``ring'' are displaying broad red-shifted
(-4 to 22.5 km s$^{-1}$) and blue-shifted velocities (-4.5 to -19.5 km s$^{-1}$), see Table 4.
Moreover, systematically the blueshifted maser spots are located to the northeast,
while the redshifted masers lie to the southwest. The velocities and size are in good agreement
with those reported by \citet{ForsterCaswell1989} and \citet{Forster1990}. However, we think
that such large radial velocity gradients ($\Delta$V $\sim$ 40 km s$^{-1}$) observed in the masers are too large to be explained
in terms of a ring in expansion produced by an energetic stellar homogenous wind that compresses
the ambient medium and drives a shock into it, and particularly since there is no a radio or infrared source. 
A possibility is that they are due 
to multiple outflows. \citet{Torrellesetal2001} and \citet{Uscangaetal2005} reported the presence
of "rings" in expansion (on much smaller scales) traced by water masers in the high mass 
star forming regions Cepheus and W75N, respectively. The radial velocities of the masers are 
close to the ambient cloud velocity.

We therefore suggest that possibly the radio sources VLA2a and VLA1a that are well centered on the positions
of the water masers might be exciting them. These radio sources thus perhaps are driving northeast-southwest
compact outflows.

\section{Summary}

We have analyzed 1.3, 2, 3.6 and 6 cm continuum and water maser line archival data from the VLA
toward the massive star forming region IRAS 17233-3606. Comparing our results with
mid-infrared images from the Spitzer Observatory's GLIMPSE survey  revealed
a cluster of young protostars associated with this region and multiple outflows,
some of them possibly related with the compact centimeter objects. We have summarized the tentative
nature of every radio source in Table 3.

The specific results and conclusions of this study are as follows.

\begin{itemize}

\item We report the detection of nine compact radio sources, six of them new detections.
      We found that they are embedded in a cluster of infrared sources associated with multiple outflows.
      Most of these radio objects appear to be ultra- and hyper-compact optically thin and thick HII regions ionized
      by early B type ZAMS stars, however, in a few cases (VLA2c and VLA2d) they could be optically thick compact 
      ($\sim$ 300 AU) and dusty circumstellar disks that might be powering a strong bow-shock and a collimated outflow 
      observed in the infrared wavelengths and that appear to emanate from this region.

\item We found that the object VLA2d is well centered in a putative low velocity, strong and compact north-south
          bipolar outflow that is traced by OH masers, and we suggest that this object would be powering it. \\

\item We suggest that the sources VLA2a and VLA1a might be energizing two northeast-southwest outflows traced
      by H$_2$O maser emission and which appear to form an ``expanding ring''.

\end{itemize}

\begin{acknowledgments}
      This research has made extensive use of the NASA's Astrophysics Data System
      and SIMBAD database operated at CDS, Strasbourg, France. We would like to thank 
      the careful referee's report which helped improving our paper.

\end{acknowledgments}

\bibliographystyle{apj}
\bibliography{zapata}
\end{document}